\documentclass[
    ,final            
  ]
  {aipproc}

\usepackage{amssymb}

\layoutstyle{6x9}

\begin{document}

\rightline{\small{DESY 07-175}}\vspace{-0.18cm}

\title{Short distance modifications to Newton's law in SUSY braneworld scenarios}


\author{Gonzalo A. Palma}{
  address={Deutsches Elektronen-Synchrotron DESY, Theory Group,
Notkestrasse 85, D-22603 Hamburg, Germany}
}

\begin{abstract}
In braneworld models coming from string theory one generally
encounters massless scalar degrees of freedom --moduli-- parameterizing the
volume of small compact extra-dimensions. Here we discuss the effects
of such moduli on Newton's law for
a fairly general 5-D supersymmetric braneworld scenario with a bulk scalar field $\phi$.
We show that the Newtonian potential describing the gravitational interaction
between two bodies localized on the visible brane picks up a non-trivial contribution at short distances
that depends on the shape of the superpotential $W(\phi)$ of the theory.
In particular, we compute this contribution for dilatonic braneworld scenarios
$W(\phi) \propto e^{\alpha \phi}$ (where $\alpha$ is a constant) and discuss
the particular case of 5-D Heterotic M-theory.
\end{abstract}

\maketitle


\section{Introduction}

In theories where matter confines to a 4-D brane
and gravity is the only massless
field able to propagate along the extra dimensional volume, one
generally expects short distance corrections to the usual 4-D Newtonian potential.
The shape and distance at which these corrections become relevant generally depend
on the geometry and size of the extra dimensional volume, thus
allowing for distinctive signals dependent of the particular content of
the theory. For instance, in the single-brane Randall-Sundrum scenario \cite{Randall:1999vf},
where a 4-D brane of constant tension $\propto k$
is immersed in an infinitely large
AdS$_{5}$ volume, a zero mode graviton $g_{\mu \nu}$ localizes about the brane.
This zero mode is exponentially suppressed away from the brane
with a warp factor $\propto e^{-kz}$, where $z$ is the
distance from the brane along the fifth extra-dimensional direction.
The Newtonian potential
describing the gravitational interaction between two bodies
of masses $m_{1}$ and $m_{2}$ in the brane, and separated
by a distance $r$, is then found to be \cite{Randall:1999vf, Garriga:1999yh, Callin:2004py}
\begin{eqnarray} \label{c1: V(r)}
V(r) = - G_{\mathrm{N}} \frac{m_{1} m_{2}}{r} \Big( 1 + \frac{2}{3 k^{2}
r^{2}} \Big), \label{V-RS}
\end{eqnarray}
where $G_{\mathrm{N}}$ is Newton's constant. The correction $2/3 k^2 r^2$ springs out directly
from the way in which gravitons propagate in an AdS$_{5}$ spacetime.
If the tension $k$ is small enough as compared to the Planck mass
$M_{\mathrm{Pl}} = (8 \pi G_{\mathrm{N}})^{-1/2}$,
then it would be possible to distinguish this type of scenario from other
extra-dimensional models in short distance tests of gravity.
Present tests \cite{Kapner:2006si, Adelberger:2006dh} give the robust constraint $1/k < 11 \mu$m.

It is therefore sensible to ask how other braneworld scenarios may differ from
the Randall-Sundrum case at short distances, especially within the context of
more realistic models. In what follows we address this question for a
fairly general class of supersymmetric braneworld scenarios with a bulk scalar field,
where the geometry of the extra-dimensional space differs from the usual AdS profile.
We show that the Newtonian potential for this type of models
picks up a non-trivial correction at scales comparable to the tension of
the brane \cite{Palma:2007tu}, that differs dramatically from the one shown in Eq. (\ref{V-RS}).

\section{SUSY Braneworlds}

\begin{figure}
  \includegraphics[height=.25\textheight]{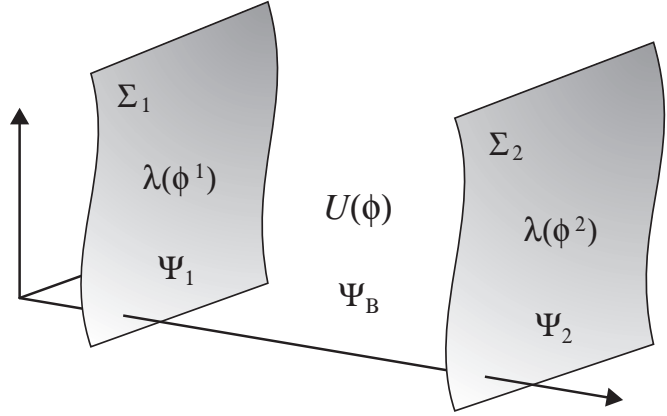}
  \caption{In the bulk there is a scalar field
$\phi$ with a bulk potential $U(\phi)$.
Additionally, the bulk space is bounded by branes $\Sigma_{1}$ and
$\Sigma_{2}$ located at the orbifold fixed points. The branes are
characterized by tensions $\lambda_{1}$ and $\lambda_{2}$, and
may contain matter fields $\Psi_{1}$ and $\Psi_{2}$ respectively.} \label{FIG}
\end{figure}

Let us consider a 5-D spacetime $M = \mathrm{R}^{4}
\times S^{1}/ \mathrm{Z}_{2}$, where $\mathrm{R}^{4}$ is a fixed 4-D
Lorentzian manifold and $S^{1}/\mathrm{Z}_{2}$ is
the orbifold constructed from a circle with points
identified through a $\mathrm{Z}_{2}$-symmetry. $M$ is bounded
by two 3-branes, $\Sigma_{1}$ and $\Sigma_{2}$, located at the fixed points of
$S^{1}/\mathrm{Z}_{2}$. There is a bulk
scalar field $\phi$ with a bulk potential $U(\phi)$ and boundary
values $\phi^{1}$ and $\phi^{2}$ at the branes. Additionally, the
branes have tensions $\lambda_{1}$ and $\lambda_{2}$ which are given
functions of $\phi^{1}$ and $\phi^{2}$
(see FIG.\ref{FIG}).
The total action of the system is
\begin{eqnarray}
S = \frac{M_{5}^{3}}{8} \int_{M} \! \big[ 4 R^{(5)}
- 3
 (\partial \phi)^{2} - 3 U(\phi) \big] - \frac{3 M_{5}^{3}}{2} \, \int_{\Sigma_{1}}
\! \lambda_{1} (\phi^{1}) - \frac{3 M_{5}^{3}}{2} \,
\int_{\Sigma_{2}} \! \lambda_{2} (\phi^{2}).
\label{c2: S-g}
\end{eqnarray}
Here $\int_{M}$ is the short notation for $\int d^{5}x \sqrt{- g_{5}}$,
where $g_{5}$ is the determinant of the 5-D metric $g_{A B}$ of signature $(-++++)$
(a similar convention follows for $\int_{\Sigma}$).
$M_{5}$ is the 5-D fundamental mass scale and $R^{(5)}$
is the 5-D Ricci scalar. Additionally,
$\lambda_{1}(\phi^{1})$ and $\lambda_{2}(\phi^{2})$ are the
brane tensions.

Our interest is focused on a class of models embedded in
supergravity, where the bulk potential $U(\phi)$ and the brane
tensions $\lambda_{1}(\phi^{1})$ and $\lambda_{2}(\phi^{2})$ satisfy
a special relation so as to preserve half of the
local supersymmetry near the branes \cite{Bergshoeff:2000zn}. This is
\begin{eqnarray}
U =  (\partial_{\phi} W)^{2} - W^{2}, \qquad
\lambda_{1} = W(\phi^{1}), \qquad \mathrm{and} \qquad
\lambda_{2} = - W(\phi^{2}), \label{BPS cond}
\end{eqnarray}
where $W = W(\phi)$ is the superpotential of the system.
Under these conditions the system presents an important property: There is
a BPS vacuum state consisting of a static bulk background
in which the branes can be allocated anywhere, without obstruction.
Indeed, suppose a metric
$ds^{2} =  dz^{2} +  g_{\mu \nu} dx^{\mu} dx^{\nu}$,
where $z$ parameterizes the extra-dimension and $g_{\mu \nu}$ is the induced
metric on the 4-D foliations of $M$ parallel to the branes.
If the bulk fields depend only on $z$ and
$g_{\mu \nu} = \omega^{2}(z) \eta_{\mu \nu}$ (with $\eta_{\mu \nu}$
the Minkowski metric) then one finds that the entire system is solved by
functions $\omega(z)$ and $\phi(z)$ satisfying
\begin{eqnarray}
\omega'/\omega = - W/ 4 \qquad \mathrm{and}
\qquad \phi' =  \partial_{\phi} W , \label{BPS-equ}
\end{eqnarray}
where $' = \partial_{z}$. Remarkably, boundary conditions at the fixed points
are also given by these two equations. Thus, the presence of the branes
forces the system to acquire a domain-wall-like vacuum background,
instead of a flat 5-D Minkowski background.

Let us mention here that in order to have the right
phenomenology in this type of scenarios, it is important
to have the moduli $\phi_{1}$ and $\phi_{2}$ stabilized (in
the 4-D effective low energy theory, they are found to be massless
and dangerously coupled to matter
\cite{Palma:2004, Palma:2005wm}). This can be done in a simple way by breaking
supersymmetry on the branes in an appropriate manner \cite{Palma:2007tu}.

\section{The Newtonian Potential}

It is possible to compute the modifications to Newton's law
arising from the way in which bulk gravitons propagate in a background
given by Eq. (\ref{BPS-equ}). In general, we can write the Newtonian potential describing
the gravitational interaction between two masses $m_{1}$ and $m_{2}$ on the 
visible brane as
$V(r) = - G_{\mathrm{N}} \frac{m_{1} m_{2}}{r} \left[ 1 + f(r) \right]$,
where $f(r)$ is a function of $r$ whose shape 
is dictated by the form of $W(\phi)$.
For concreteness, let us consider the case of dilatonic braneworlds
$W(\phi) = \Lambda \, e^{\alpha \phi}$ where
$\Lambda > 0$ is some fundamental mass scale. Then, 
the background geometry of the system is
\begin{eqnarray}
\phi(z) = \phi_{1} - \frac{1}{\alpha} \ln \left[1 - \alpha^{2} W_{0} z \right], \qquad
\mathrm{and} \qquad
\omega(z) = \left[ 1 - \alpha^{2} W_{0} z \right]^{1/4\alpha^{2}}.
\end{eqnarray}
Notice the presence of a singularity $\omega=0$ at $z = 1/ \alpha^{2} W_{0}$.
Without loss of generality, one may take the position of
$\Sigma_{1}$ at $z=0$ (since $\Lambda>0$, this is a positive tension brane). Then,
$\Sigma_{2}$ can be anywhere between $z=0$ and
$z = 1/ \alpha^{2} W_{0}$. Interestingly, the relevant case of
5-D Heterotic M-theory \cite{Horava:1996ma, Lukas:1998tt}
corresponds to $\alpha^{2} = 3/2$.

If, for simplicity, we further assume that the visible brane is $\Sigma_{1}$
while the second brane
$\Sigma_{2}$ is very close to the bulk singularity, then it is possible to find
three different solutions for $f(r)$, depending on the value of $\alpha$
(see \cite{Palma:2007tu} for a detailed analysis of this)
\begin{eqnarray}
f(r) =  \left\{
\begin{array}{ll}
 \frac{8}{3\pi^{2}} \frac{1 - 4\alpha^{2}}{1+2\alpha^{2}}
\int_{0}^{\infty} \frac{dm}{m}
\frac{e^{-mr}}{ J_{\nu - 1}^{2} [m/k] + Y_{\nu - 1}^{2}[m/k]} & \qquad \textrm{if $\alpha^{2} < 1/4$} \\
\frac{32}{9 \pi W_{0}}  \int_{\sqrt{b}}^{\infty} dm
\frac{\sqrt{m^{2} - b}}{m} e^{-mr} & \qquad \textrm{if $\alpha^{2} = 1/4$} \\
\frac{4}{3} \frac{4 \alpha^{2} -1}{1 + 2 \alpha^{2}}
\sum_{n} e^{- k \, u^{\mu+1}_{n} r} & \qquad \textrm{if $\alpha^{2} > 1/4$}
\end{array} \right.
\end{eqnarray}
In the previous expression we have defined $b =  (3 W_{0}/8)^{2}$,
$k \equiv |1 - 4 \alpha^{2}| W_{0}/4$, 
$\nu \equiv \frac{3}{2} (1 - 4 \alpha^{2})^{-1} + \frac{1}{2}$
and $\mu \equiv \frac{3}{2} (4 \alpha^{2} -1)^{-1} - \frac{1}{2}$, 
where $W_{0} = \Lambda \, e^{\alpha \phi_{1}}$
with $\phi_{1}$ the value of $\phi$ at the positive tension brane.
Additionally, $u^{\mu+1}_{n}$ is the $n$-th zero 
of the Bessel function $J_{\mu+1}[x]$,
that is $J_{\mu+1}[u^{\mu+1}_{n}] = 0$. Observe that
Eq. (\ref{V-RS}) is recovered for $\alpha = 0$.

\section{Discussion}

We have shown the corrections to Newton's law for SUSY braneworld models arising
from the way in which gravitons propagate in a bulk with a geometry that differs
from the more commonly studied AdS$_{5}$.
A sensible question regarding this type of models is whether there are any chances of
observing short distance modifications of general relativity in the near future.
To explore this, notice that the relevant energy scale at which the corrections become
significant is $W_{0} = \Lambda \, e^{\alpha \phi_{1}}$, instead of the
more fundamental mass scale $\Lambda$. Typically one would expect
$\Lambda \simeq M_{5}$ which has to be above TeV scales to agree with
particle physics constraints.
Nevertheless, the factor
$e^{\alpha \phi_{1}}$ leaves open the possibility
of bringing $\lambda = W_{0}^{-1}$ up to micron scales.

In the case of 5-D Heterotic M-theory ($\alpha^{2} = 3/2$) one has
$e^{\alpha \phi_{1}} = 1/\mathcal{V}$,
where $\mathcal{V}$ is the
volume of the Calabi-Yau 3-fold in units of $M_{5}$.
In order to have an accessible scale $\lambda \simeq 10 \mu$m,
it would be required
$\mathcal{V} \, \frac{M_{\mathrm{Pl}}}{M_{5}} \simeq 10^{29}$,
where we assumed $\Lambda \simeq M_{5}$.
On the other hand, Newton's constant is given by
$G_{\mathrm{N}}^{-1} = \frac{32 \pi}{1+2\alpha^{2}} M_{5}^2 W_{0}^{-1}$, which
implies $M_{\mathrm{Pl}}^{2} \simeq M_{5}^{2} \mathcal{V}$. Thus, to
achieve $\mathcal{V} \, \frac{M_{\mathrm{Pl}}}{M_{5}} \simeq 10^{29}$ one
requires the following values for $M_{5}$ and $\mathcal{V}$
\begin{eqnarray}
M_{5} \simeq 10^{-10} M_{\mathrm{Pl}}, \qquad \mathrm{and} \qquad
\mathcal{V}^{1/6} \simeq 10^{3},
\end{eqnarray}
which are in
no conflict with present phenomenological constraints coming from
high energy physics. In particular, non-zero Kaluza-Klein modes coming from
the compactified volume $\mathcal{V}$ would have masses of order $10^{6}$GeV.
On the other hand, if $M_{5}$ is of the order of the grand unification scale
$M_{\mathrm{GUT}} \sim 10^{16}$GeV, then corrections to the Newtonian potential
would be present at the non-accessible scale $\lambda \sim 10^{-20} \mu$m.


\begin{theacknowledgments}
This work was supported by the Collaborative Research Centre 676
(Sonderforschungsbereich 676) Hamburg, Germany.
\end{theacknowledgments}



\bibliographystyle{aipproc}   





\end{document}